\documentclass{article}
\usepackage{spconf,amsmath,graphicx,hyperref}
\usepackage{booktabs}
\usepackage{xcolor}
\usepackage{colortbl}
\usepackage{amsfonts}
\usepackage[subtle]{savetrees}

\title{A Parameter-Efficient Multi-Scale Convolutional Adapter for Synthetic Speech Detection}
\name{Yassine El Kheir$^{1,3}$,
      Fabian Ritter-Guttierez$^{4}$,
      Arnab Das$^{1,3}$,
      Tim Polzehl$^{1,3}$,
      Sebastian Möller$^{1,2}$}

\address{$^{1}$Speech and Language Technology, DFKI, Germany\\
$^{2}$Quality and Usability Lab, Technical University of Berlin, Germany\\
$^{3}$AI Team, Gretchen AI, Germany\\
$^{4}$Nanyang Technological University, Singapore
}

\begin{document}
\ninept
\maketitle
\begin{abstract}
%Adapting large pre-trained SSL models for synthetic speech detection presents a critical trade-off between performance and efficiency. Full fine-tuning is resource-intensive, while current parameter-efficient methods often fail to model the complex, multi-level artifacts found in spoofed audio. This paper introduces MultiConvAdapter, a novel adapter architecture that resolves this trade-off. We propose injecting multi-scale convolutional modules into the attention layers of SSL encoder. This allows the model to learn discriminative features at multiple temporal resolutions simultaneously, effectively capturing both short-term artifacts and long-term distortions. With only 3M trainable parameters, MultiConvAdapter significantly reduces the computational burden of adaptation. On five public datasets, including ASVspoof, MLAAD and In-The-Wild, MultiConvAdapter demonstrates superior performance over traditional fine-tuning and competing adapter methods.

Recent synthetic speech detection models typically adapt a pre-trained SSL model via finetuning, which is computationally demanding. Parameter-Efficient Fine-Tuning (PEFT) offers an alternative. However, existing methods lack the specific inductive biases required to model the multi-scale temporal artifacts characteristic of spoofed audio. This paper introduces the Multi-Scale Convolutional Adapter (MultiConvAdapter), a parameter-efficient architecture designed to address this limitation. MultiConvAdapter integrates parallel convolutional modules within the SSL encoder, facilitating the simultaneous learning of discriminative features across multiple temporal resolutions, capturing both short-term artifacts and long-term distortions. With only $3.17$M trainable parameters ($1\%$ of the SSL backbone), MultiConvAdapter substantially reduces the computational burden of adaptation. Evaluations on five public datasets, demonstrate that MultiConvAdapter achieves superior performance compared to full fine-tuning and established PEFT methods.

%. This paper present MultiConvAdapter, a parameter efficient architecture that learns multi-level artifacts while using less number of parameters compared to popular PEF methods such as LoRa. MultiConvAdapter learns multiple temporal resolution in parallel, capturing short-term artifacts and long-term distortions. With only 3M trainable parameters, MultiConvAdapter significantly reduces the computational burden of adaptation. On five public datasets, including ASVspoof, MLAAD and In-The-Wild, MultiConvAdapter demonstrates superior performance over traditional fine-tuning and competing adapter methods such as LoRa.

\end{abstract}
\section{Introduction}

Recent improvements in generative AI, specifically text-to-speech (TTS) and voice conversion (VC) now produces speech that is nearly
%Recent advances in generative AI have dramatically improved speech synthesis. State-of-the-art text-to-speech (TTS) and voice conversion (VC) systems now produce speech that is nearly 
indistinguishable from natural human utterances \cite{hao2025boosting, rubenstein2023audiopalm}.%The integration of large language models (LLMs) into speech generation pipelines \cite{hao2025boosting, zhang2023speechgpt, rubenstein2023audiopalm} has further enhanced the fidelity and expressiveness of synthetic speech.
While these developments are beneficial, they as well pose risks. The growing accessibility of voice generation technologies has fueled their misuse to spread misinformation
%, hate speech, and even terrorism-related content 
\cite{meaker2023deepfake}. Hence, there is a current importance for deploying a reliable system for distinguishing between genuine and spoofed speech.

% Moreover, synthetic speech is increasingly exploited in financial fraud, with reports showing that 7.7\% of individuals have been affected, including high-profile cases such as a CEO in the United Kingdom targeted by voice cloning scams \cite{mcafee2023beware}.

%These threats underscore the urgent need for robust countermeasure systems capable of reliably distinguishing between genuine and spoofed speech.

Early approaches to synthetic speech detection primarily relied on manually engineered spectral features such as LFCC \cite{chen2021ur}, MFCC \cite{caceres2021biometric}, CQCC \cite{das2021known}, Short-Time Fourier Transform (STFT) and Mel-spectrograms \cite{muller2022does}. These features capture high-frequency artifacts and are typically paired with classifiers such as CNNs or MLPs for binary discrimination between bonafide and spoofed speech. Various CNN-based architectures have been investigated \cite{muller2022does, dong2023multi,wang2021comparative,wang2024multi}.
%Various CNN-based architectures have been investigated, including ResNet \cite{muller2022does}, Res2Net \cite{dong2023multi}, ECAPA-TDNN \cite{wang2021comparative}, and LCNN \cite{wang2024multi}.

Parallel efforts have explored end-to-end systems that operate directly on raw audio waveforms \cite{ge2021raw}. Notably, the AASIST model \cite{jung2022aasist} combines fixed sinc-convolutional filter banks with a spectro-temporal graph neural network (GNN) equipped with attention mechanisms, achieving stronger generalization than earlier handcrafted pipelines.

Recently, the field has shifted towards leveraging Self-Supervised Learning (SSL) models, including Wav2Vec2.0 \cite{xie2023learning}, XLSR \cite{wang2024can}, wavLM and HuBERT \cite{yang2024robust} as audio feature extractor backbone. These models, pre-trained on large-scale unlabeled data, provide robust representations that generally outperform traditional methods, particularly under unseen attack conditions \cite{kheir2025comprehensive}. However, the deployment of SSL models in downstream tasks like synthetic speech detection often requires full fine-tuning, which is computationally expensive and prone to overfitting when task-specific data is limited \cite{muller2024harder}.

A common alternative to full-finetuning is Parameter-Efficient Fine-Tuning (PEFT), where a lightweight adapter module is used,  
%A promising direction is to adapt pre-trained SSL encoders through lightweight adapter modules, 
enabling efficient fine-tuning while keeping most parameters frozen. Adapters have already shown to be able to match or outperform full-finetuning in the NLP and computer vision domain \cite{hu2022lora, houlsby2019parameter, karimi2021compacter}. However, their direct application to synthetic speech detection is not straightforward. Existing designs struggle to capture the multi-level spoofing cues, which are manifested across multiple temporal scales, that characterize speech signals \cite{wang2023low}. These cues are further distorted by channel noise, codec artifacts, and compression effects. More recent efforts, such as local–global adapter architectures \cite{wu2024adapter}, have achieved satisfactory results but at the cost of over $50$M additional parameters, which is nearly 16\% of the total SSL model size. Thus, such approach would need an additional $50$M parameters during deployment.

\label{sec:intro}

\begin{figure}
    \centering
    \includegraphics[width=1\linewidth]{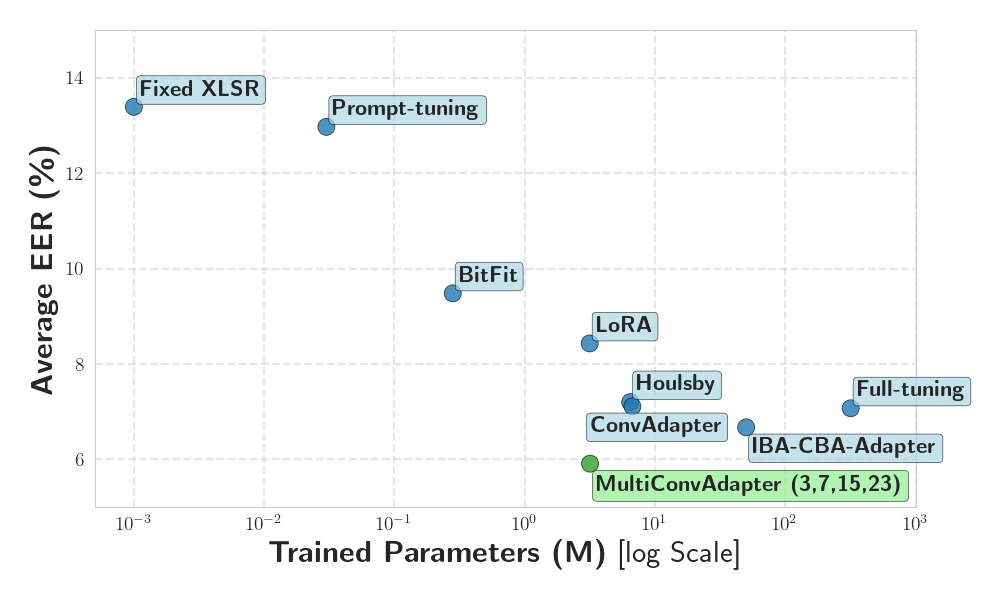}
    \vspace{-6mm}
    \caption{Trainable parameters vs EER\% trade-off for PEFT methods. }
    \label{fig:tardeoff}
    \vspace{-4mm}
\end{figure}

\begin{figure*}
    \centering
    \includegraphics[width=0.95\linewidth]{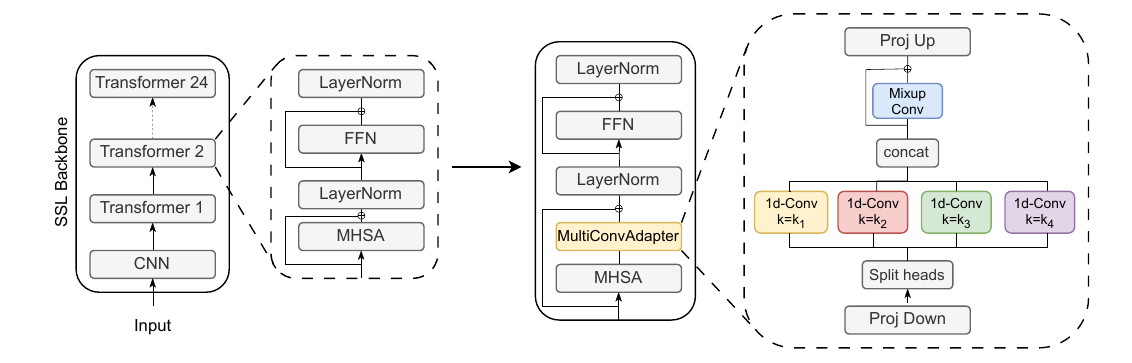}
    \vspace{-4mm}
    \caption{Our proposed MultiConvAdapter, Multi-Scale Convolutional Adapters with multi-scale kernels $\{k_{1}, k_{2}, k_{3}, k_{4}\}$ for effecient synthetic speech detection}
    \label{fig:MultiConvAdapter}
    \vspace{-4mm}
\end{figure*}
\vspace{-0.3mm}

Crucially, spoofed speech contains both short-term artifacts and long-term distortions introduced by synthesis algorithms and real-world transmission conditions. An effective anti-spoofing system must be sensitive to both. To this end, we propose MultiConvAdapter, a parameter-efficient adapter that augments SSL backbones with only 3.17M trainable parameters (1\% of the backbone). By introducing multi-scale convolutions after the self-attention layers, the model becomes more sensitive to both short-term artifacts and long-term distortions in spoofed speech. As shown in Fig.~\ref{fig:tardeoff}, our approach achieves a favorable trade-off between accuracy and efficiency: it yields a 16.41\% relative reduction in EER compared to full fine-tuning, and a 29.9\% relative improvement over LoRA while using a similar number of parameters (3.17M vs. 3.15M). Extensive ablation studies confirm the importance of adapter placement, and consistent gains across five public datasets and multiple SSL backbones (XLSR, HuBERT, WavLM) highlight the robustness of the proposed method. Overall, MultiConvAdapter provides a simple and effective way to strengthen speech anti-spoofing systems while remaining highly parameter-efficient.

\section{Methodology}
\subsection{System Overview.}

We propose to add our lightweight MultiConvAdapter that learns inductive bias on different temporal resolutions within the Transformer layers of an SSL backbone, as depicted in Fig. \ref{fig:MultiConvAdapter}. The principle is to augment the standard transformer architecture of the SSL backbone with a parallel processing path specifically designed to capture the multi-scale temporal artifacts characteristic of synthetic speech while using as fewer parameters as possible.  As shown in Fig. \ref{fig:MultiConvAdapter}, the MultiConvAdapter module is injected into each transformer layer, operating on the output of the Multi-Head Self-Attention (MHSA) sub-module. This placement allows the adapter to leverage the global contextual information from the self-attention mechanism while introducing a strong inductive bias for localized temporal patterns.

\subsection{MultiConvAdapter: Multi-Scale Convolutional Adapter}

The design of MultiConvAdapter is motivated by a core principle from state-of-the-art speech architectures: complementing the global context from self-attention with the strong local inductive bias of convolutions, as established by Conformer \cite{gulati2020conformer}. 

% Furthermore, recent work, MultiConvFormer \cite{prabhu2024multi}, has validated that extending this to a multi-scale convolutional design is critical for handling the complexities of speech.

The MultiConvAdapter module, detailed in Fig. \ref{fig:MultiConvAdapter} (right), is a bottleneck architecture that utilizes parallel depthwise convolutions to maintain a low parameter count. Let 
$H_l \in \mathbb{R}^{B \times T \times D}
$ be the activations from the MHSA at layer $l$, with $B$ the batch size, $T$ the feature sequence length and $D$ the embedding dimension. While MHSA excels at capturing global contextual information, it lacks a mechanism to explicitly capture multiple temporal artifact granularities. 

Standard PEFT methods, such as LoRA, introduce an algebraic inductive bias by assuming adaptation occurs within a low-rank subspace. However, they do not inherently enforce structural priors related to the temporal nature of audio data. Our proposed MultiConvAdapter aims to explicitly model artifacts at different temporal granularities, complementing the global context captured by the MHSA mechanism. MultiConvAdapter takes the activations $H_l$ and projects it to a lower-dimensional space $H'_l = W_{down} * H_l$ to reduce computational complexity. The lower-dimensional projection is done across the $H_l$ feature dimension $D$. Hence, $H'_l \in \mathbb{R}^{B \times T \times D'}$ for $D' << D$.

The projected features $H'_l$ are divided along the channel dimension ($D'$) into $N$ heads. Each head of dimension $D'//N$ is then processed by a separate 1D depthwise convolution with a distinct kernel size $k_i$. The use of depthwise convolutions is crucial not only for efficiency but also for modelling purposes. Using a depthwise convolution allows for temporal modelling to be decoupled from the channel modelling. This design enables the parallel extraction of features at multiple temporal resolutions, with smaller kernels capturing local, high-frequency artifacts and larger kernels modeling longer-term distortions, while each of them is done at each channel separately.

After this parallel multi-scale feature processing, the outputs from all heads are concatenated back to form a representation of dimension $D'$, as before. A fusion module, referred to as \textit{\textbf{Mixup Conv}}, is then applied to integrate information across the branches (Fig. \ref{fig:MultiConvAdapter}). This module consists of a residual 1D-convolution with a kernel size of 3, which enables effective information exchange among multi-scale features extracted. The fusion step allows the model to learn an optimal combination of multi-scale representations for the downstream task. Finally, the fused vector is projected back to dimension $D$.

\begin{table*}[h!]
% \setlength\tabcolsep{0.91pt}
% \renewcommwand{\arraystretch}{0.91}
\centering
\caption{Comparison of different parameter-efficient fine-tuning strategies across multiple datasets. Values represent mean EER (\%) over three experiments. \textbf{Parameters} refers to the total number of trainable parameters. The \textbf{AVG} column reports the average EER across all test sets. \dag: System results reproduced. }
\label{tab:peft_comparison}
\scalebox{0.9}{
\begin{tabular}{l|c|@{\hspace{0.5cm}}ccccc@{\hspace{0.5cm}}|c}
\toprule
\textbf{} & \textbf{Parameters} & \textbf{LA19} & \textbf{DF21} & \textbf{ITW} &  \textbf{MLAAD} & \textbf{ASV5} & \textbf{AVG} \\
\midrule
Full-tuning     & $317$ M & 0.35 & 2.60 & 9.43 & 15.86 & 7.12 & 7.07 \\
\midrule
Fixed XLSR           & $0$ M & 1.52 & 7.02 & 29.70 & 17.03 & 11.69 & 13.39 \\
Prompt-tuning\textsuperscript{\dag} \cite{Oiso_2024}    & $0.03$ M & 1.79 & 5.90 & 24.71 & 20.97 & 11.50 & 12.97 \\
BitFit (Bias-tuning) \cite{zaken2021bitfit}            & $0.28$ M & 0.79 & 3.06 & 13.64 & 18.53 & 11.37 & 9.48 \\
LoRA\textsuperscript{\dag} \cite{laakkonen2025generalizable} & $3.15$ M& 0.61 & 4.33 & 13.15 & 16.85 & 7.22 & 8.43 \\
Houlsby \cite{houlsby2019parameter} & $6.44$ M& 0.58 & 2.88 & 10.55 & 15.57 & 6.42 & 7.20 \\
ConvAdapter \cite{li2023evaluating} & $6.70$ M& 0.67 & 2.23 & 9.81 & 16.70 & 6.13 & 7.11 \\
IBA-CBA-Adapter\textsuperscript{\dag} \cite{wu2024adapter}        & $50$ M & 0.42 & 2.07 & 8.47 & 15.86 & 6.52 & 6.67 \\

\midrule
\textbf{Our Proposed} & & & & & &\\
MultiConvAdapter (\(\{3,7,15,23\}\))             &  $3.17$ M& 0.56 & 1.89 & \cellcolor{green!20}\textbf{7.92} & \cellcolor{green!20}\textbf{13.23} & \cellcolor{green!20}\textbf{5.97} & \cellcolor{green!20}\textbf{5.91} \\

MultiConvAdapter (\(\{7,15,23,31\}\))             &  $3.17$ M& \cellcolor{green!20}\textbf{0.27} & \cellcolor{green!20}\textbf{1.75} & 8.29 & 17.31 & 6.41 & 6.81 \\

\bottomrule

\end{tabular}}
\vspace{-2mm}
\end{table*}

% The outputs from the N depthwise convolution branches are first concatenated along the channel dimension. This concatenated feature map then passes through a fusion block, denoted as "Mixup Conv" in Fig. 1. This block, implemented as a 1x1 grouped convolution with a residual connection, facilitates information exchange between the features extracted at different temporal scales. This fusion step allows the model to learn the optimal combination of multi-scale features for the downstream task. Finally, the fused representation is projected back to the original embedding dimension D using a 1x1 convolution. The entire MultiConvAdapter module operates within a residual connection, adding its output to the initial input activations $H_l$, ensuring that the adapter learns additive, fine-grained refinements to the features from the SSL backbone.

\label{sec:methodology}

\section{Experimental Design}

\subsection{Dataset and evaluation metrics} \label{subsec:datasets}
We evaluate our approach on five benchmark datasets, ASVspoof LA (LA19) \cite{wang2020asvspoof}, ASVspoof2021 DF (DF21) \cite{yamagishi2021asvspoof}, In-The-Wild (ITW) \cite{muller2022does}, MLAAD \cite{muller2024mlaad}, and ASVspoof5 (ASV5) \cite{wang2024asvspoof5}. These corpora reflect diverse spoofing conditions, from canonical TTS/VC-generated attacks to large-scale, adversarial, and in-the-wild samples, and are widely used to assess robustness and cross-domain generalization. Specifically, (i) \textbf{DF21} contains deepfake-style attacks with lossy compression simulating realistic storage and transmission, with $14{,}869$ bonafide and $519{,}059$ spoofed utterances; (ii) \textbf{ITW} consists of $31{,}779$ audio clips collected from online sources (e.g., podcasts, political talks), totaling approx.\ $20.7$ h bonafide and $17.2$ h spoofed; (iii) \textbf{MLAAD}, the English subset of MLAAD v3, includes $37{,}998$ samples equally distributed between bonafide and spoofed, generated with state-of-the-art open-source TTS and VC systems; and (iv) \textbf{ASV5}, the latest large-scale benchmark, comprises $138{,}688$ genuine and $542{,}086$ fake utterances covering $32$ spoofing techniques, including adversarial samples.

For fair comparison, we use the standard LA19 train–dev split for training and validation, and evaluate on the LA19, DF21, ITW, and MLAAD test sets. In addition, we run experiments on ASV5 using its official train–test partitions to further assess performance on the latest large-scale and adversarial benchmark.

We evaluate model performance using the Equal Error Rate (EER), the standard metric in anti-spoofing tasks.

\subsection{SSL backbone and classifier architecture}

% We employ a pre-trained XLSR-53 model as the feature extraction backbone . This model is based on the Wav2Vec 2.0 architecture and consists of a multi-layer convolutional feature encoder that processes the raw waveform, followed by a 24-layer transformer encoder..... 

We employ a pre-trained Wav2Vec2.0 XLSR-53 (XLSR) as it has demonstrated strong capabilities in speech deepfake detection \cite{kheir2025comprehensive}. XLSR consists of an encoder network of $7$ blocks of temporal convolution layers with $512$ channels, and a context network of $24$ transformer blocks with model dimension $1024$, inner dimension $4096$, and $16$ attention heads. We use the widely-adopted AASIST~\cite{jung2022aasist} as the back-end classifier.

\subsection{Implementation Details}

Audio samples are truncated or padded to a fixed length of 4 seconds (\(64{,}600\) samples). Models are trained with the Adam optimizer~\cite{diederik2014adam} using \(\beta_1=0.9\) and \(\beta_2=0.999\). We train all models for 50 epochs with a batch size of 14, a learning rate of \(1\times10^{-5}\), and a weight decay of \(1\times10^{-4}\). Cross-entropy loss is used as the objective. To enhance data diversity, we apply noise injection, reverberation, and SpecAugment. Experiments were conducted on a single NVIDIA H100 GPU. Reported results are averaged over three runs with different random seeds.
%All training runs are executed on a single NVIDIA H100 GPU, and each experiment is repeated three times with different random seeds to account for variability. Each reported result accounts for the mean of the three seeds.

The MultiConvAdapter uses \(D' = 64\) with convolutional kernel sizes \(\{3,7,15,23\}\). LoRA follows the setup in \cite{wang2023low} with rank $16$. The Houlsby adapter employs a projection dimension of $64$ to ensure a fair comparison with our proposed adapter.

% Audio samples are truncated or
% padded to a fixed length of approximately 4 seconds (64,600). Model training is performed using the Adam optimizer~\cite{diederik2014adam} with $\beta_1 = 0.9$ and $\beta_2 = 0.999$. All models are trained for 50 epochs with a batch size of 14, with a learning rate of \(1 \times 10^{-5}\) and a weight decay of \(1 \times 10^{-4}\). The cross-entropy loss is used as the objective function. All training runs are executed on a single NVIDIA H100 GPU. To ensure robustness and account for variability, each experiment is repeated three times using different random seeds.

% MultiConvAdapter, $D'$ is set to 64, and kernels by default are {3,7,15,23}. for Lora, we followed \cite{wang2023low} with a rank of 16. and for Houslby we opt for a projection dimension of 64 (for fair comparsion with our proposed adapter).

\section{Results and Analysis}
% \vspace{-6mm}

\begin{table}[h!]
\caption{Performance comparison of different kernel configurations across multiple datasets.}
\label{tab:kernel_comparison}
\centering
\scalebox{0.83}{
\begin{tabular}{l|@{\hspace{0.25cm}}ccccc@{\hspace{0.25cm}}|c}
\toprule
\textbf{Kernels} & \textbf{LA19} & \textbf{DF21} & \textbf{ITW} &  \textbf{MLAAD} & \textbf{ASV5} & \textbf{AVG} \\
\midrule
\multicolumn{7}{c}{\textit{No kernels}} \\
\midrule
$ \emptyset $    & 0.58 & 2.46 & 10.49 & 17.05 & 6.51 & 7.42 \\
\midrule
\multicolumn{7}{c}{\textit{Single-head}} \\
\midrule
$\{3\}$     & 0.49 & 2.10 & 8.46 & 16.79 & 6.88 & 6.94 \\
$\{7\}$     & 0.45 & 1.98 & 8.77 & 17.42 & 6.50 & 7.02 \\
$\{15\}$    & 0.45 & 1.94 & \textbf{7.71} & 16.65 & 6.55 & 6.66 \\
$\{23\}$    & 0.34 & 2.84 & 10.64 & 17.06 & 6.38 & 7.45 \\
$\{31\}$    & 0.47 & 2.05 & 8.47 & 18.12 & 6.33 & 7.09 \\
\midrule
\multicolumn{7}{c}{\textit{Two-heads}} \\
\midrule
$\{3,15\}$   & 0.31 & 2.39 & 9.17 & 18.19 & 6.07 & 7.23 \\
$\{7,23\}$   & 0.42 & 1.96 & 8.55 & 17.07 & 6.49 & 6.90 \\
$\{3,23\}$   & 0.45 & 1.97 & 8.40 & 16.05 & \textbf{5.92} & 6.56 \\
$\{15,23\}$   & 0.34 & 1.86 & 8.14 & 16.60 & 6.01 & 6.59 \\
\midrule
\multicolumn{7}{c}{\textit{Four-heads}} \\
\midrule
$\{3,7,15,23\}$  & 0.56 & 1.89 & 7.92 & \textbf{13.23} & 5.97 & \textbf{5.91} \\
$\{7,15,23,31\}$ & \textbf{0.27} & \textbf{1.75} & 8.29 & 17.31 & 6.41 & 6.81 \\
\bottomrule
\end{tabular}}
\vspace{-4mm}
\end{table}

% In this section, we first evaluate the performance of our proposed MultiConvAdapter against full fine-tuning and several established PEFT methods. Next, we analyze the impact of different kernel configurations to validate our multi-scale design hypothesis. Finally, we conduct a comprehensive ablation study to assess the contributions of key architectural components.

\subsection{How does MultiConvAdapter compares with PEFT Strategies}

Table \ref{tab:peft_comparison} compares the performance and parameter efficiency of different adaptation strategies across the datasets described in Section \ref{subsec:datasets}. As it can be seen, MultiConvAdapter is consistently superior across all the datasets evaluated. Specifically the $\{3, 7, 15, 23\}$ configuration achieves the lowest average EER of $5.91\%$, representing a relative reduction of 16.41\% in average compared to the full-finetuning, despite using only $3.17$~M trainable parameters, only $1\%$ of the $317$~M parameters required for full fine-tuning. This result shows the effectiveness of our approach in achieving competitive or even superior performance while drastically reducing computational requirements.

%As presented in Table~\ref{tab:peft_comparison}, our proposed MultiConvAdapter demonstrates a superior balance of parameter efficiency and detection performance compared to existing methods. Most notably, MultiConvAdapter achieves a significantly lower average EER of $5.91\%$ compared to the full fine-tuning baseline, despite using only $3.17$~M trainable parameters, only $1\%$ of the $317$~M parameters required for full fine-tuning. This result shows the effectiveness of our approach in achieving competitive or even superior performance while drastically reducing computational requirements.

Compared with other PEFT methods, MultiConvAdapter consistently outperforms them. It achieves a lower average EER than methods with similar parameter counts, such as LoRA ($8.43\%$), Houlsby ($7.20\%$) and ConvAdapter ($7.11\%$), and exceeds methods with fewer parameters, such as Prompt-tuning ($12.97\%$) and BitFit ($9.48\%$). Even when compared to the much larger IBA-CBA-Adapter \cite{wu2024adapter}, which uses $50$~M trainable parameters, our compact $3.17$~M parameter model delivers a better average performance ($5.91\%$ vs. $6.67\%$). LoRA imposes an algebraic inductive bias by assuming adaptation occurs within a low-rank subspace \cite{hu2022lora}. Houlsby adapters utilize MLP-based bottlenecks \cite{houlsby2019parameter}. Neither method explicitly enforces structural priors related to the temporal dynamics of audio signals. Although ConvAdapter and IBA-CBA-Adapter use convolutions, they are restricted to a single, local kernel size. This design is insufficient for capturing the full range of spoofing artifacts, which explains the consistently better performance of MultiConvAdapter's multi-scale approach.

% ConvAdapter and IBA-CBA-Adapter, though using convolutions, are restricted to a single local kernel, which is insufficient for capturing the full range of spoofing artifacts which may explain the consistently better performance of MultiConvAdapter.

MultiConvAdapter introduces a strong structural inductive bias through multi-scale convolutions, enabling localized feature learning and weight sharing across time. This specialized structure is inherently better suited to modeling these temporal artifacts compared to generic adaptation mechanisms. Results in Table \ref{tab:peft_comparison} also highlight the adaptability of our multi-kernel configurations. The $(\{3,7,15,23\})$ setup shows exceptional generalization, achieving the best performance on three out of five datasets, particularly on challenging sets ITW, MLAAD, and ASV5. In contrast, the configuration with larger kernels $(\{7,15,23,31\})$ excels in the LA19 and DF21 datasets.%,, suggesting its proficiency in capturing artifacts specific to these challenges. 

\vspace{-2mm}
\subsection{Impact of Kernel Configurations}

To validate our hypothesis on the importance of multi-scale artifacts modeling for spoof detection, we analyzed the performance of various kernel configurations, with results detailed in Table~\ref{tab:kernel_comparison}. The baseline model with no convolutional kernels ($\emptyset$), equivalent to a standard Houlsby adapter on MHSA, yields $7.42\%$ average EER. The introduction of even a single convolutional head provides a notable improvement, with the $\{15\}$ kernel achieving the best single-head performance.

% However, no single kernel configuration is uniformly best across all datasets. LA19 and DF21 tend to improve as kernel size increases, a trend visible in configurations that emphasize larger receptive fields (e.g., $\{15,23\}$ spans $\approx$ $300-460$ ms, and $\{7,15,23,31\}$ spans $\approx$ $140–620$ ms), implying that distinguishing cues in those sets often extend over several hundred milliseconds which aligns with previous findings on LA19, DF21 \cite{das2025generalizable}. By contrast, ITW is best served by a single mid-range kernel (the $\{15\}$ head). While MLAAD acted well with a kernel combing small and large receptive field: {3, 23}. In short, datasets differ in the temporal extent of their salient artifacts, so a single configuration kernel can be optimal for some datasets but suboptimal for others.

However, no single kernel configuration is uniformly best across all datasets. LA19 and DF21 tend to benefit from larger kernels: configurations that emphasize larger receptive fields (e.g., $\{15,23\}$$\approx$$300–460$ms; $\{7,15,23,31\}$$\approx$$140–620$ms) yield lower EERs, suggesting that distinguishing cues in these sets manifest over several hundred milliseconds, consistent with prior findings \cite{das2025generalizable}. By contrast, ITW is best served by a single mid-range kernel ($\{15\}$$\approx$300ms). MLAAD responds well to a combination spanning short and long contexts ($\{3,23\}$$\approx$$60–460$ms). Datasets differ in the temporal extent of their salient artifacts: a single fixed kernel can be optimal for some datasets but suboptimal for others.

The multi-scale (multi-head) approach resolves this dependency; four-head configurations deliver the best overall trade-off between per-dataset specialization and cross-dataset robustness. The hierarchical set $(\{3,7,15,23\})$ attains the lowest average EER (5.91\%) while remaining robust on the most challenging evaluations (MLAAD: 13.23\%, ASV5: 5.97\%) and still achieving strong results on LA19, DF21, and ITW. Thus, using all four kernels simultaneously effectively removes the need for dataset-specific kernel selection, the multi-kernel adapter captures short, mid, and long temporal artifacts, and yields the best generalization across heterogeneous attack types and recording conditions.

% However, multi-head configurations prove optimal across all datasets, indicating that different spoofing attacks exhibit artifacts at varying temporal resolutions.

% Combining kernels consistently leads to better and more robust performance. The four-head configurations deliver the best results, confirming the benefits of a hierarchical receptive field. The $(\{3,7,15,23\})$ configuration achieves the lowest overall average EER of $5.91\%$ while being robust on the most challenging datasets (MLAAD: $13.23\%$, ASV5: $5.97\%$). Meanwhile, the $(\{7,15,23,31\})$ configuration, which favors larger kernels, shows a strong affinity for the LA19 and DF21 datasets. These findings strongly suggest that the ability to simultaneously process audio features at multiple time scales is a key factor in the success of the MultiConvAdapter, enabling it to generalize more effectively across diverse spoofing techniques.

\begin{table}[h!]
\vspace{-4mm}
\centering
\caption{Ablation on feature aggregation and position of the adapter.}
\label{tab:ablation_study}
\scalebox{0.85}{
\centering
\begin{tabular}{l|@{\hspace{0.25cm}}ccccc@{\hspace{0.25cm}}|c}
\toprule
\textbf{Setting} & \textbf{LA19} & \textbf{DF21} & \textbf{ITW} & \textbf{MLAAD} & \textbf{ASV5} & \textbf{AVG} \\
\midrule
% Fusion \\ (Mixup Conv) & 0.56 & 1.89 & \cellcolor{green!20}\textbf{7.92} & \cellcolor{green!20}\textbf{13.23} & \cellcolor{green!20}\textbf{5.97} & \cellcolor{green!20}\textbf{5.91} \\

\shortstack{Fusion \\ (Mixup Conv)} & 0.56 & 1.89 & \textbf{7.92} & \textbf{13.23} & \textbf{5.97} & \textbf{5.91} \\

%\midrule
%\multicolumn{7}{c}{\textit{Dimension ($D$)}} \\
%\midrule
%$D'=32$  & 0.78 & 2.15 & 8.76 & 17.25 & 6.11 & 7.01 \\
%$D'=128$ & 0.43 & 1.93 & 8.45 & 13.76 & 5.98 & 6.11 \\
\midrule
\multicolumn{7}{c}{\textit{Aggregation Strategy}} \\
\midrule
Sum & \textbf{0.41} & 2.23 & 9.44 & 18.30 & 6.14 & 7.30 \\
Concat & 0.46 & 2.03 & 9.36 & 17.48 & 6.25 & 7.12 \\
Weighted Sum & 0.43 & \textbf{1.86} & 8.27 & 16.49 & 6.77 & 6.76 \\
\midrule
\multicolumn{7}{c}{\textit{Position of Adapter}} \\
\midrule
FFN        & 0.54 & 2.56 & 9.82 & 15.75 & 6.52 & 7.04 \\
MHSA and FFN & 0.47 & 2.45 & 8.92 & 14.09 & 6.01 & 6.39 \\
\bottomrule
\end{tabular}}
\vspace{-6mm}
\end{table}

\subsection{Ablation on channel-aggregation strategy and adapter position}

We conducted an ablation study, summarized in Table~\ref{tab:ablation_study}, to investigate the aggregation strategy of features, and the position of the adapter within the transformer block.

Our default fusion approach, using Mixup Conv, outperforms alternative strategies such as Weighted Sum, Concatenation (Concat), and Summation (Sum). Both Concat and Sum led to a notable drop in performance, yielding average EERs of $7.12\%$ and $7.30\%$, respectively. 

Notably, simple strategies (Sum / Weighted Sum) can provide an advantage on in-domain or near-domain sets that share recording distributions: LA19 and DF21. However, Mixup Conv substantially outperforms these alternatives on out-of-domain evaluations (ITW, MLAAD, ASV5), demonstrating better generalization to unseen conditions. This improvement stems from Mixup Conv’s 1D convolutional fusion, which models explicit cross-scale interactions instead of treating multi-scale features as independent. This mechanism allows the model to learn discriminative patterns based on how short-term artifacts co-occur with long-term distortions, providing a much richer and more context-aware representation.

% This performance gap suggests that simply combining the multi-scale features as if they are independent is insufficient. The Mixup Conv fusion approach is more effective because it uses a 1D convolution to explicitly model the interactions between the features from different scales. This mechanism allows the model to learn discriminative patterns based on how short-term artifacts co-occur with long-term distortions, providing a much richer and more context-aware representation.

% Sum, Concat and Weighted Sum assumes that the features extracted by different kernels are directly additive. However, artifacts may exhibit non-linear dependencies. On the other hand, our Mixup Conv does not imposes such strong constraint which may explain the performance improvement.

The second analysis investigates the placement of MultiConv Adapter within the transformer block. Results suggest that the most effective strategy is positioning the adapter immediately after the MHSA module. This placement likely allows the adapter's specialized local analysis to directly leverage the globally-aware representations generated by the attention mechanism. Other placements, while still yielding strong results, were found to be less optimal. In contrast, positioning the adapter only after the Feed-Forward Network (FFN) yielded a higher EER of $7.04\%$, suggesting the adapter is less effective after the FFN's point-wise feature transformations have been applied. the FFN has been optimized during the SSL pre-training objective, which may project the features into a representation space where the localized temporal artifacts for spoof detection are more difficult to model. Placing an adapter after both MHSA and FFN achieved a competitive EER of $6.39\%$, reinforcing the primary importance of the post-MHSA placement for our architecture.

% Finally, we assessed the optimal placement of the adapter module. Inserting the adapter after both the MHSA and MLP layers results in a superior average EER of $6.39\%$ compared to placing it only after the FFN layer ($7.04\%$). The rationale for this observation relates to the distinct functional roles of the transformer components and how they complement the convolutional adapter. Placing the MultiConvAdapter immediately after the MHSA allows the module to inject localized, multi-scale temporal information directly into the globally attended features. This interleaving of global (attention) and local (convolutional adapter) processing creates a fused representation that incorporates both global context and local hierarchy.

% \begin{table}[h!]
% \label{tab:sota_table}
% \centering
% \caption{Comparison with the SOTA Single Systems on ASV19}
% \scalebox{0.8}{
% \centering
% \begin{tabular}{l|ccccc}
% \toprule
% \textbf{Setting} & \textbf{LA19} & \textbf{DF21} & \textbf{ITW} & \textbf{MLAAD} & \textbf{ASV5} \\
% \midrule

% MultiConvAdapter & 0.56 & 1.89 & \cellcolor{green!20}\textbf{7.92} & \cellcolor{green!20}\textbf{13.23} &  \\

% \bottomrule
% \end{tabular}}
% \end{table}

\begin{table}[h!]
\vspace{-4mm}
\centering
\caption{Comparison of Full-tuning vs MultiConvAdapter using different SSL backbones. Relative improvement from Full-tuning to MultiConvAdapter is shown ($\downarrow$ indicates relative reduction).}
\label{tab:full_vs_adapter}
\scalebox{0.75}{
\begin{tabular}{l|ccccc|c|c}
\toprule
\textbf{Model} & \textbf{LA19} & \textbf{DF21} & \textbf{ITW} & \textbf{MLAAD} & \textbf{ASV5} & \textbf{AVG} & \textbf{Rel. Imp.} \\
\midrule
\multicolumn{8}{c}{\textit{XLSR}} \\
\midrule
Full-tuning      & \textbf{0.35} & 2.60 & 9.43 & 15.86 & 7.12 & 7.07 & - \\
MultiConvAdapter & 0.56 & \textbf{1.89} & \textbf{7.92} & \textbf{13.23} & \textbf{5.97} & \textbf{5.91} & $\downarrow$16.41\% \\
\midrule
\multicolumn{8}{c}{\textit{HuBERT}} \\
\midrule
Full-tuning      & \textbf{0.54} & \textbf{5.28} & 20.72 & 24.72 & 14.81 & 13.21 & - \\
MultiConvAdapter & 0.83 & 5.62 & \textbf{16.61} & \textbf{22.24} & \textbf{9.65} & \textbf{10.99} & $\downarrow$16.80\% \\
\midrule
\multicolumn{8}{c}{\textit{WavLM}} \\
\midrule
Full-tuning      & 0.64 & 3.76 & 18.18 & 15.88 & 8.17 & 9.33 & - \\
MultiConvAdapter & \textbf{0.55} & \textbf{2.80} & \textbf{14.81} & \textbf{13.82} & \textbf{5.27} & \textbf{7.45} & $\downarrow$20.15\% \\
\bottomrule
\end{tabular}}
\vspace{-6mm}
\end{table}

\subsection{Generalization Across SSL Backbones}

To verify that the benefits of our proposed adapter are not limited to a single SSL backbone, we compared MultiConvAdapter against full fine-tuning on three different SSL backbones: XLSR, HuBERT, and WavLM. As shown in Table \ref{tab:full_vs_adapter}, MultiConvAdapter consistently provides a significant performance advantage. MultiConvAdapter achieves a relative improvement in average EER of 16.41\% with XLSR, 16.80\% with HuBERT\footnote{https://huggingface.co/facebook/hubert-large-ll60k}, and 20.15\% with WavLM\footnote{https://huggingface.co/microsoft/wavlm-large} over their fully fine-tuned counterparts. These results demonstrate that MultiConvAdapter is a robust, model-agnostic strategy.

\begin{table}[h!]
\vspace{-4mm}
\centering
\caption{Comparison of Full-tuning vs MultiConvAdapter across different classifiers. Relative improvement from Full-tuning to MultiConvAdapter is shown ($\downarrow$ indicates relative reduction).}
\label{tab:classifiers_full_vs_adapter}
\scalebox{0.75}{
\begin{tabular}{l|ccccc|c|c}
\toprule
\textbf{Model} & \textbf{LA19} & \textbf{DF21} & \textbf{ITW} & \textbf{MLAAD} & \textbf{ASV5} & \textbf{AVG} & \textbf{Rel. Imp.} \\
\midrule
\multicolumn{8}{c}{\textit{AASIST}} \\
\midrule
Full-tuning      & \textbf{0.35} & 2.60 & 9.43 & 15.86 & 7.12 & 7.07 & - \\
MultiConvAdapter & 0.56 & \textbf{1.89} & \textbf{7.92} & \textbf{13.23} & \textbf{5.97} & \textbf{5.91} & $\downarrow$16.41\% \\
\midrule
\multicolumn{8}{c}{\textit{BiCrossMamba-ST}} \\
\midrule
Full-tuning      & \textbf{0.46} & 2.22 & 10.35 & 12.73 & 6.01 & 6.35 & - \\
MultiConvAdapter & \textbf{0.46} & \textbf{1.82} & \textbf{8.22} & \textbf{10.58} & \textbf{6.34} & \textbf{5.48} & $\downarrow$13.70\% \\
\midrule
\multicolumn{8}{c}{\textit{Nes2Net}} \\
\midrule
Full-tuning      & 0.45 & 3.41 & 10.95 & 11.40 & 6.31 & 6.50 & - \\
MultiConvAdapter & \textbf{0.39} & \textbf{2.15} & \textbf{7.15} & \textbf{11.40} & \textbf{5.82} & \textbf{5.38} & $\downarrow$17.23\% \\
\bottomrule
\end{tabular}}
\vspace{-6mm}
\end{table}

\subsection{Generalization Across Classifiers}

To examine whether the effectiveness of our proposed adapter extends beyond a specific classifier architecture, we evaluate MultiConvAdapter against full fine-tuning across three representative classifiers: AASIST, BiCrossMamba-ST \cite{kheir2025bicrossmamba}, and Nes2Net \cite{liu2025nes2net}. As shown in Table~\ref{tab:classifiers_full_vs_adapter}, MultiConvAdapter consistently yields lower average EER compared to full-tuning across all classifiers. Specifically, it achieves a relative improvement of 16.41\% with AASIST, 13.70\% with BiCrossMamba-ST, and 17.23\% with Res2Net. These results confirm that MultiConvAdapter generalizes well across diverse classifier designs, highlighting its flexibility and strong regularization effect regardless of the underlying architecture.

\section{Conclusions}

This paper introduced the Multi-Scale Convolutional Adapter (MultiConvAdapter), a parameter-efficient approach for adapting pre-trained SSL models to the task of synthetic speech detection. MultiConvAdapter addresses the multi-temporal resolution nature of synthetic speech by integrating parallel depthwise convolutions with varying kernel sizes directly into the transformer layers of the SSL backbone. Such design allows the model to maintain the global modelling capabilities of the SSL backbone while also learning varying temporal artifacts from our proposed MultiConvAdapter. Experiments across five public datasets demonstrated the efficacy of this approach, obtaining the best average performance compared to other PEFT methods while adding only 1\% of trainable parameters. Ablation studies validates our architectural choices, in particular the placement of the adapter after the MHSA module as well as the Mixup Conv for fusion of the multi-temporal learned representations. Future work will explore dynamic kernel selection mechanisms. 

\bibliographystyle{IEEEbib}
\bibliography{strings,refs}

\end{document}